# Anomalous electrochemical capacitance in Mott−insulator titanium sesquioxide

Sumana Kumar[1†], Sukanta Nandi[1†], Vikash Mishra[2], Alok Shukla[2], Abha Misra[1‡]

[1]Department of Instrumentation and Applied Physics, Indian Institute of Science, Bangalore, Karnataka 560012, India.
[2]Department of Physics, Indian Institute of Technology Bombay, Powai, Mumbai, Maharashtra 400076, India.

**Abstract**

Electrochemical capacitors with pure electric double layer capacitance are so far largely been limited to carbon materials only. Conventional metal oxides with Faradaic pseudocapacitance substantially suffer from the material instability at high temperatures and thus there is a demand of novel metal oxides exhibiting thermally improved high electric double layer capacitance with material stability. In this study, titanium sesquioxide, $Ti_2O_3$, a Mott−insulator in its metallic state at 300 °C has shown an anomalous increase of ~560% in the solid−state electrochemical capacitance as compared to its pristine response at room temperature (RT). In aqueous electrolyte, the maximum capacitance was obtained up to 1053 $\mu F/cm^2$, which is 307% higher than with solid−state electrolyte. Moreover, the semiconducting state of $Ti_2O_3$ demonstrated ~3500% enhancement in its electrochemical capacitance upon infrared illumination at RT. The observed electrochemical responses in $Ti_2O_3$ are attributed to the transition of semiconducting to the metallic state by redistribution of the localized electrons. Our experimental results find a good agreement with the first−principles density−functional theory calculations that revealed an increase in the charge−density with the rise in temperature, which is mainly attributed to the surface Ti atoms. The study opens wide avenues to engineer the electrochemical double layer capacitance of $Ti_2O_3$ with high chemical stability.

[†] Both the authors have equal contribution.
[‡] Corresponding author, Email: abha.misra1@gmail.com

**Introduction**

Fundamental limitations in batteries, like solid−state diffusion rate, phase transformations and the change in volume during charge/discharge have led to the preference of electrochemical capacitors, from the small−scale systems to large cells, for high rate energy transfer applications. Conway described the general requirement for the electrochemical capacitor materials as oxides with high electrical conductivity and exhibiting two or more oxidation states without a change in the phase.[1] First such widely studied materials are $RuO_2$, $IrO_2$ and $MnO_2$ with pseudo−capacitive properties, where nonstoichiometry is mainly responsible for the electronic conductivity.[1] Moreover, electrochemical properties of these oxide films are highly dependent on their thickness, deposition temperature and mostly successful with aqueous electrolytes, thereby limiting their applications in the planar devices. The development of planar micro−scale electrochemical capacitor devices has established a direct connection with wearable smart devices covering a wide range of applications like health, environmental, industrial and transportation.[2] Thus, the quest for novel materials continues for a large electrochemical double layer capacitance and compatible with the planar micro−scale electrochemical capacitor devices.

Titanium oxides, the most abundant materials exhibit mixed valence states of Ti ions. Among these, titanium sesquioxide ($Ti_2O_3$) with $Ti^{3+}$ ions and a narrow direct bandgap ($E_g$ ~0.11 eV) has been studied recently for its photothermal characteristics[3] and long wavelength mid−infrared detection.[4] $Ti_2O_3$ also known as a Mott−insulator attains a metallic conductivity at around 450 K without any change in its phase.[5] Vanadium oxides, another class of Mott−insulators with identical $d^1$ configuration are reported for their electrochemical capacitive behaviour in many works; however, poor stability severely affects their utility due to the lower transition temperature induced by the structural transition.[6,7] Therefore, higher transition temperature in $Ti_2O_3$ without structural transition is important for the devices operable under high-temperature conditions.

We report $Ti_2O_3$ based micro−electrochemical capacitor in a planar geometry as a novel electrode material. The granular $Ti_2O_3$ is directly spray-coated on a planar interdigitated electrode surface. Solid−state micro−electrochemical capacitance in the symmetric configuration is evaluated after annealing $Ti_2O_3$ in a wide temperature range from room temperature (RT) to 400 °C. An increase in areal electrochemical capacitance is observed in annealed $Ti_2O_3$ with a maximum enhancement of ~560% after annealing at 300 °C. Moreover, an optical interaction in the infrared (IR) range enhances the electrochemical capacitance to ~3500% in the pristine $Ti_2O_3$. This anomalous increase in the electrochemical capacitance has not been reported so far in any metal oxide materials used for electrochemical capacitance thus, suggesting two distinct mechanisms contributing to electrochemical response in the Mott−insulator, $Ti_2O_3$. A high capacitance retention of 80% in $Ti_2O_3$ device is achieved after 5000 cycles. Furthermore, the $Ti_2O_3$ device annealed at 300 °C in aqueous electrolyte exhibits areal capacitance of ~1050 μF/cm² at a scan rate of 1 mV/s. Our work demonstrated a promising strategy to rationally design and fabricate novel $Ti_2O_3$ material−based devices with largely enhanced capacitive behaviour.

**Experimental details**

The granular powder of $Ti_2O_3$ was commercially procured from Sigma−Aldrich with particle size of 100 mesh as inferred from the scanning electron microscope (SEM) micrographs (for an example Figure S1). Dispersion of 200 mg $Ti_2O_3$ is prepared in 30 ml of ethanol by ultrasonication for 6 hours. The electrochemical performance of Ti$_2$O$_3$ was analysed using planar interdigitated gold electrodes deposited on a glass substrate with 20/200 nm thick chromium/gold (Cr/Au) layers for both the anode and the cathode. Each electrode consists of four fingers and the dimension of each finger is 1x13 mm$^2$ with an active electrode area of 2.4 cm$^2$. The mass loading on microelecteodes was 2.2 mg/cm$^2$. The dispersion of Ti$_2$O$_3$ was spray coated on both the electrodes and prior to the electrochemical measurements, the spray coated electrodes were annealed at 200 °C, 300 °C and 400 °C for 6 hours.

**Results and Discussion**

Detailed characterizations were conducted to elucidate the effect of annealing on the granular Ti$_2$O$_3$. The X−ray diffraction (XRD) of Ti$_2$O$_3$ confirmed the corundum crystal structure due to the presence of corresponding crystalline planes (101), (012), (104), (110), (006), (113), (202), (024), (116), (122), (214), (300), (1010), (220), (306) and (312) at 2θ = 20.71°, 23.82°, 33.05°, 34.82°, 39.67°, 40.23°, 43.02°, 48.70°, 53.78°, 56.21°, 61.33°, 62.44°, 72.43°, 73.54°, 76.58° and 78.61°, respectively at all the thermal conditions (figure 1(a) and figure S2).[3,8,9] The remaining peaks indicated by * are identified as other thermally stable TiO$_x$ phases and do not reveal any variations due to the annealing at any temperature.[10] The lattice parameters *a*, *b* and *c* of a trigonal cell in the $R\bar{3}c$ corundum crystal of Ti$_2$O$_3$ deduced from XRD analysis at RT are *a=b*=5.146 Å and *c*=13.636 Å. The parameters *c, a* and the ratio *c/a* at different annealing temperatures are shown in figure 1(b). The change in *c* and *a* parameters of Ti$_2$O$_3$ with annealing temperature is shown to follow the linear and nonlinear variations, respectively in the upper figure in 1(b). The value of *a* changes marginally from 5.146 (RT) to 5.145 Å (400 °C), while that of *c* changes considerably from 13.636 (RT) to 13.643 Å (400 °C). A large variation in *c* leads to a higher change in the ratio *c/a*, being 2.652 at 400 °C (the lower figure in 1(b)) compared to other annealing conditions. Moreover, in agreement with the previous studies,[5] no additional peaks are observed in Ti$_2$O$_3$ after annealing at 200 °C and 300 °C (figure 1(c)), thus confirming no impurity phases in $Ti_2O_3$. However, at 400 °C, a peak corresponding to (110) plane of rutile titanium dioxide ($TiO_2$) appeared at 2θ=27.48°, as shown in figure 1(d). Furthermore, a close inspection of (300) plane is depicted in figure 1(e−h) from RT to 400 °C. There is no peak corresponding to (002) plane of rutile $TiO_2$ at both RT and 200 °C as depicted in Figures 1(e) and (f). However, at 300 °C, a secondary peak corresponding to (002) plane of rutile $TiO_2$ appears as shown by an arrow in figure 1(g), which became prominent at 400 °C. Figure 1(h) prominently depicts a peak at 62.38° for (002) plane of rutile $TiO_2$ at 400 °C along with (300) plane of $Ti_2O_3$ at 62.48°.[11]

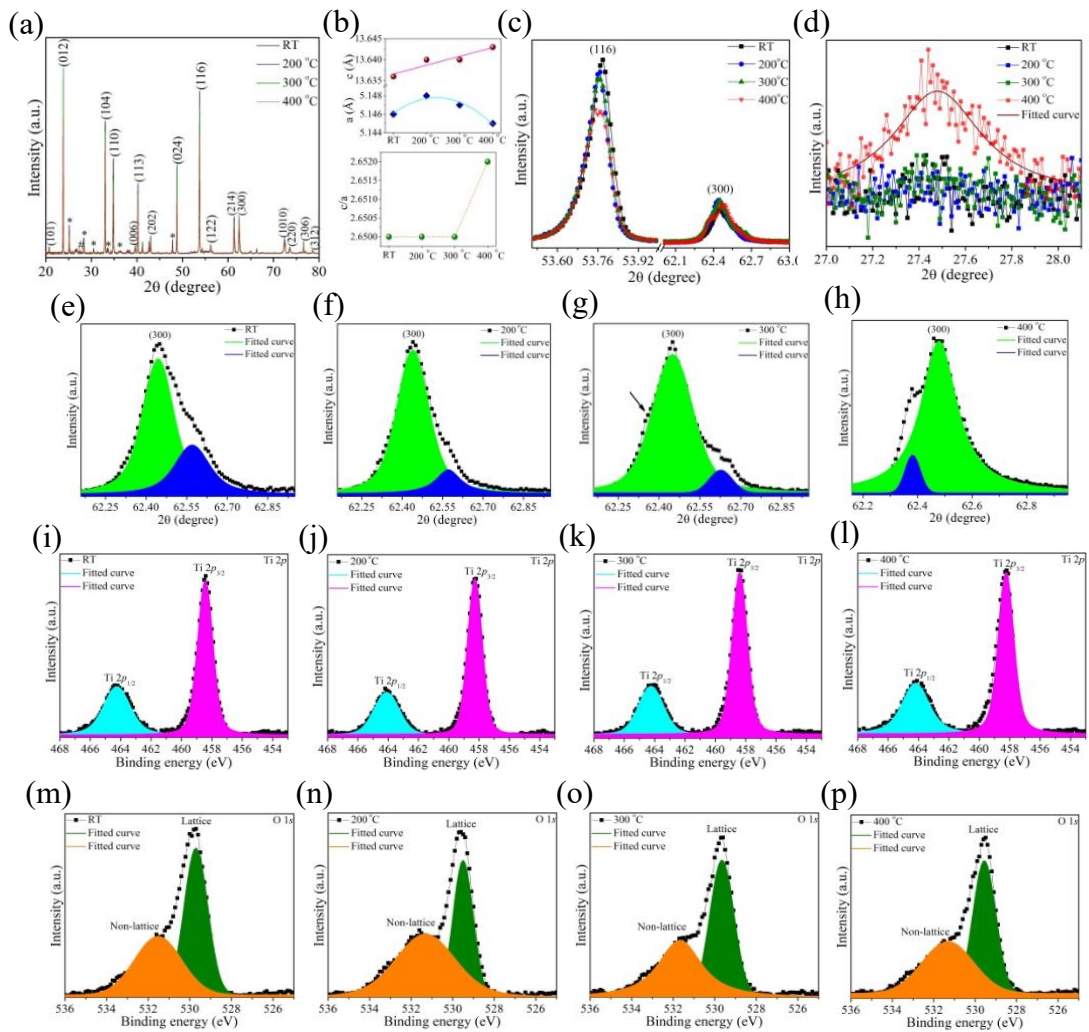

Figure 1. (a) XRD spectra of pristine $Ti_2O_3$ and after annealing at 200 °C, 300 °C and 400 °C. (b) Lattice parameters *c* and *a* evaluated from the XRD analysis are plotted with the annealing temperature (top and bottom, respectively). (c) (116) and (300) planes of $Ti_2O_3$ at heating

conditions. (d) Zoomed−in XRD spectra showing emergence of (110) plane of rutile $TiO_2$ in annealed $Ti_2O_3$ at 400 °C and compared with other heating conditions. (e)−(h) A systematic modification is depicted alongside (300) plane of $Ti_2O_3$ at both RT and annealed temperatures. XPS spectra of pristine $Ti_2O_3$ at RT and annealed at 200 °C, 300 °C and 400 °C to show the electronic states respectively of Ti in (i−l) and the corresponding spectra for O in (m−p).

Electronic states of the constituent elements (Ti and O) of $Ti_2O_3$ at different heating conditions are probed with X−ray photoelectron spectroscopy (XPS) as depicted in figure 1(i−p). It is observed in figure 1(i) that Ti is in +3 state with the spin−orbit splitting resulting in the electronic states of Ti $2p_{3/2}$ (458.4 eV) and Ti $2p_{1/2}$ (464.2 eV).[4,8] The peak positions are monitored in different annealed samples without any noticeable differences *e.g.* the peak for Ti $2p_{3/2}$ at RT shifts from 458.4 eV (figure 1(i)) to 458.2 eV at 400 °C (figure 1(l)). The oxygen on the other hand is in O *1s* state, with the peak position at 529.7 eV (figure 1(m)) owes its origin from Ti−O bonds in the lattice of $Ti_2O_3$.[4,8] In addition, a secondary peak (non−lattice) at 531.5 eV (figure 1(m)) is also observed that could be originated from the distorted $Ti_2O_3$ structure.[5,8] Similar to the Ti peak shifts, there are no observable peak shift noted with O at the different annealing temperatures e.g. the lattice peak of O at 529.7 eV at RT (figure 1(m)) only changes to 529.6 eV at 400 °C (figure 1(p)). The peak positions at all the annealing temperatures are provided in figure S3(a) for the Ti and figure S3(b) for the O spectra.

The SEM image in figure 2(a) depicts the surface of the pristine $Ti_2O_3$ and a change in the surface morphology after annealing at 400 °C is clearly revealed in figure 2(b), with nanostructures as suggested by Li *et al.* in their report.[5] The change in surface morphology can be attributed to the phase transformation of $Ti_2O_3$ to rutile $TiO_2$ with a $Ti_2O_3/TiO_2$ core−shell structure.[5,12] However, the samples annealed at 200 °C and 300 °C do not show any observable morphological differences on the surface (figure S4).

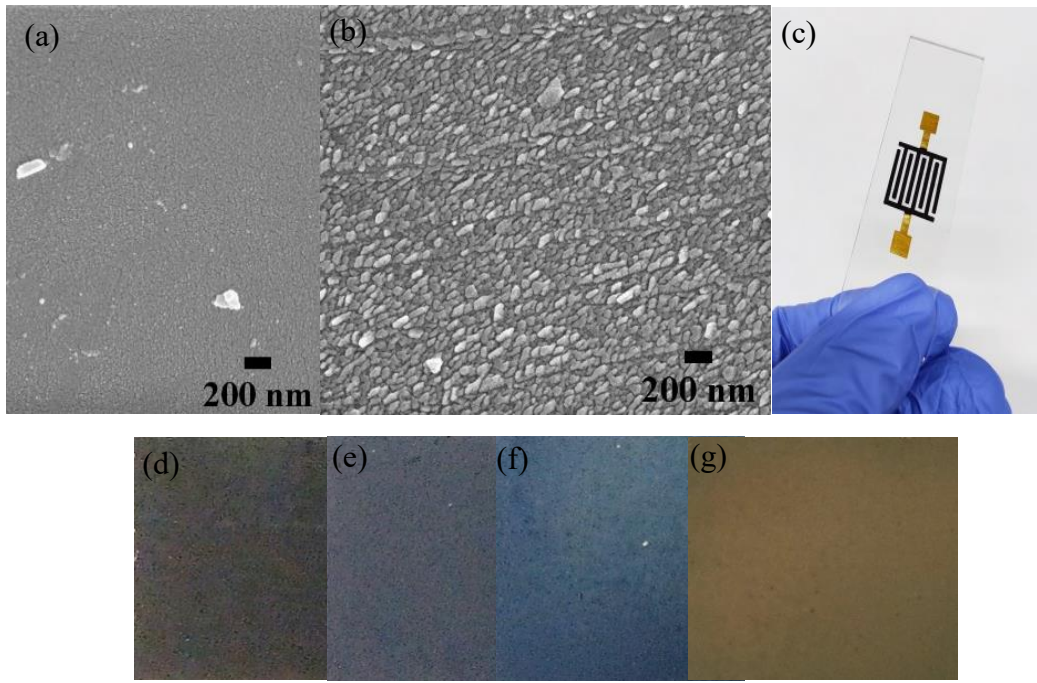

Figure 2. SEM images of (a) pristine $Ti_2O_3$ and after annealing at 400 °C (b). (c) Digital image of the spray−coated interdigitated planar device. Optical images (at same magnification) are showing a change in the colour of pristine $Ti_2O_3$ (d) at RT along with the annealing temperature at (e) 200 °C, (f) 300 °C, and (g) 400 °C.

The optical images in figure 2(d−g) show a change in colour of the surface of pristine $Ti_2O_3$ (figure 2(d)) after annealing at 200 °C (figure 2(e)), 300 °C (figure 2(f)) and 400 °C (figure 2(g)). A drastic change in colour is observed at an annealing temperature of 400 °C compared to other annealing conditions for $Ti_2O_3$. These results agree with both the XRD measurements and a report by Li *et al.*, where the formation of $Ti^{4+}$ layer on the surface of $Ti_2O_3$ is reported even at RT, which becomes prominent after annealing.[5] Energy dispersive X−ray analysis (figure S5) revealed a decrease in the intensities of both Ti and O at 400 °C. An apparent surface modification in $Ti_2O_3$ is revealed in both the optical and SEM images, which is supported by the XRD and XPS measurements, thus showing only surface modifications without any dominant new phase due to the annealing of $Ti_2O_3$.

Further, density functional theoretical (DFT) calculations were performed to understand the impact of annealing on electronic transition in $Ti_2O_3$. Details of simulation parameters are provided in the supporting information. All electronic properties of $Ti_2O_3$ were calculated using lattice parameters derived from the temperature−dependent XRD data. Figure 3(a) shows an indirect band diagram of $Ti_2O_3$ at RT with a bandgap of 0.11 eV (at $\Gamma-Z$ points), which is shown to decrease with the increase in the temperature.

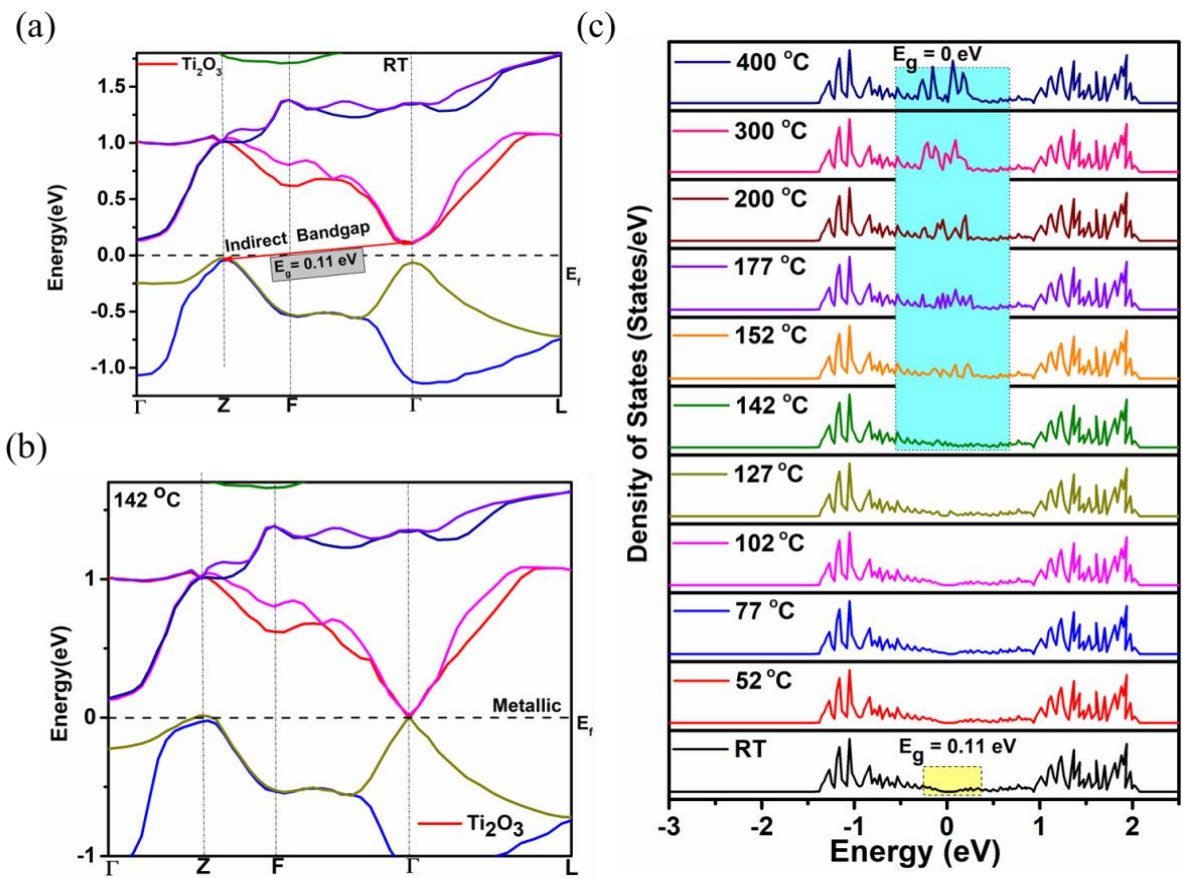

Figure 3. Band structure of $Ti_2O_3$ at (a) RT (b) at 142 °C. (c) Total DOS of $Ti_2O_3$ at different temperatures and shows the presence of bandgap in the range from RT to 142 °C.

A band merging ($E_g = 0$) is observed at around 142 °C and thus a transition to a metallic state of $Ti_2O_3$ can be seen at $\Gamma$ point (figure 3(b)).[13] The HSE06[14] calculations were performed for the total density of states (TDOS) of $Ti_2O_3$ to evaluate a modification in $E_g$ with the temperature variation in the range from RT to 400 °C (figure 3(c)).

It is observed that the $E_g$ decreases gradually with an increase in the temperature and a complete metallic transition is observed at a temperature of ~142 °C, which is attributed to the emergence of new states between the conduction band maxima (CBM) and valance band minima (VBM). Figure S6(a) clearly shows a variation in bandgap with temperature and a metallic transition around 142 °C in $Ti_2O_3$. Figure S6(b) shows the partial density of states (PDOS) at RT, where the electrons of Ti−Ti bonding are highly populated near VBM and CBM. It is important to note that the metallic behaviour in $Ti_2O_3$ is mainly due to the hoping of the $3d$ electrons in the Ti−Ti bond. The electronic phase transition therefore arises from a delicate balance between the width and energy position of Ti−$3d$ sub−bands along with the variation in Ti−Ti bond length. At RT, the states near Fermi level ($E_f$) are dominated by Ti−$3d$ orbitals (figure S6(b)), suggesting a strong hybridization[15,16] between of Ti−Ti bonding due to $3d$ orbitals.[13] Furthermore, a closer inspection of the density of states near $E_f$ (at higher temperature (see figure S6(c)) revealed a dominant participation of Ti−$3d$ orbitals in the formation of Ti−Ti bonding than Ti−O bond arising due to the interaction with O−$2p$ orbitals. This further confirms the finding in XPS results. The Bader charge analysis has also been carried out in the VASP CHGCAR using charge densities grid and shown in Table S1 & S2. The charge analysis further reveals an increase in the charge−density with the rise in temperature, which is mainly attributed to the surface Ti atoms. On the contrary, the charge−density remains unchanged for the bulk Ti atoms (i.e. inner atoms) and all the O atoms.

The increase in charge density on the surface of $Ti_2O_3$ is further exploited to study the electrochemical capacitance due to the formation of electric double layer on the surface of the electrodes. Figure 2(c) depicts the planar interdigitated electrodes after spray coating of pristine $Ti_2O_3$ in a symmetric configuration. The micro−electrochemical capacitor is fabricated by

drop−casting PVA/H3PO4 (poly(vinyl alcohol)/phosphoric acid) solid gel electrolyte. The gel was prepared by the method described elsewhere.[17] Both the electrodes were directly probed with the CHI600E electrochemical station. The electrochemical performance of Ti$_2$O$_3$ devices annealed at different temperatures was measured using cyclic voltammetry (CV) at various scan rates ranging from 5 mV/s to 1000 mV/s (figure 4(a)−(d)). A rectangular shape of CV without redox peaks is observed in both the pristine as well as annealed Ti$_2$O$_3$ based electrochemical capacitors. Therefore, it clearly indicates the presence of a large capacitive response at the interface of electrode material surface and electrolyte.

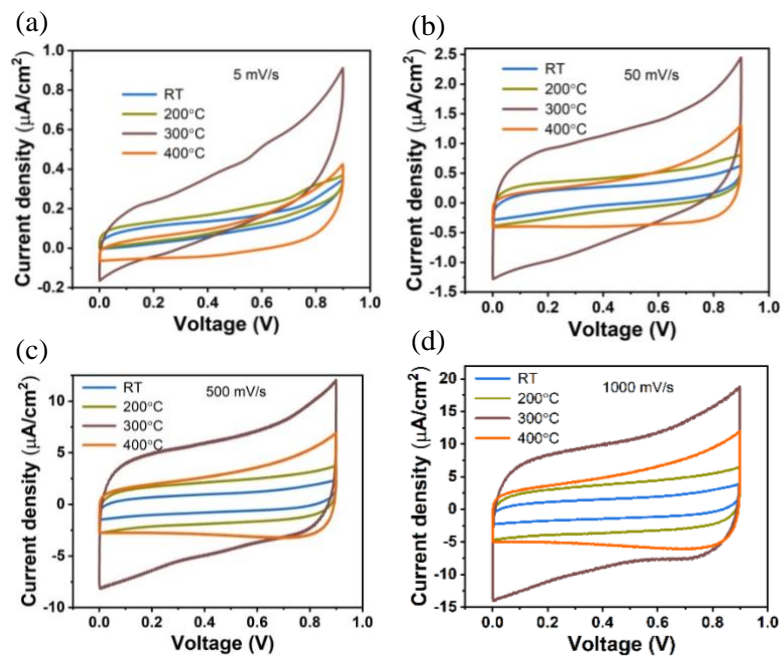

Figure 4. CV curves of the planar interdigitated electrochemical capacitor evaluated with pristine and annealed Ti$_2$O$_3$ devices at the scan rate of (a) 5 mV/s, (b) 50 mV/s, (c) 500 mV/s, and 1000 mV/s.

The rectangular shape of CV at a high scan rate further indicates a low internal resistance and high rate capability of the capacitor. Therefore, the electrochemical behaviour of Ti$_2$O$_3$ is unconventionally different from materials like RuO$_2$ and MnO$_2$, where distorted CV curves were obtained due to the presence of higher pseudocapacitance. Moreover, a high−temperature electrochemical stability of $Ti_2O_3$ is attributed to its high structural stability till 300 °C. Total charge storage in the $Ti_2O_3$ devices can be distinctly deconvoluted into two components: electrical double layer capacitance (EDLC) and diffusion-controlled processes (pseudocapacitance) using Dunn's method.[18][19] In the CV curve, the current response (i) at a fixed potential (V) can be expressed as the sum of the two separate mechanisms, namely surface capacitive effects ($k_1 v$) and diffusion-controlled process ($k_2 v^{1/2}$) according to the following relation:

$$i(V) = k_1 v + k_2 v^{1/2} \qquad (1)$$

or $$\frac{i(V)}{v^{1/2}} = k_1 v^{1/2} + k_2 \qquad (2)$$

where $v$ is the scan rate.[18] By determining $k_1$ and $k_2$ in equation 2, a separate contribution is evaluated from the current arising from diffusion-controlled process and surface capacitive process at specific potentials. Figure 5(a)-(d) exhibit the capacitive and diffusion-controlled contribution in pristine and annealed $Ti_2O_3$ devices at a scan rate of 100 mV/s.

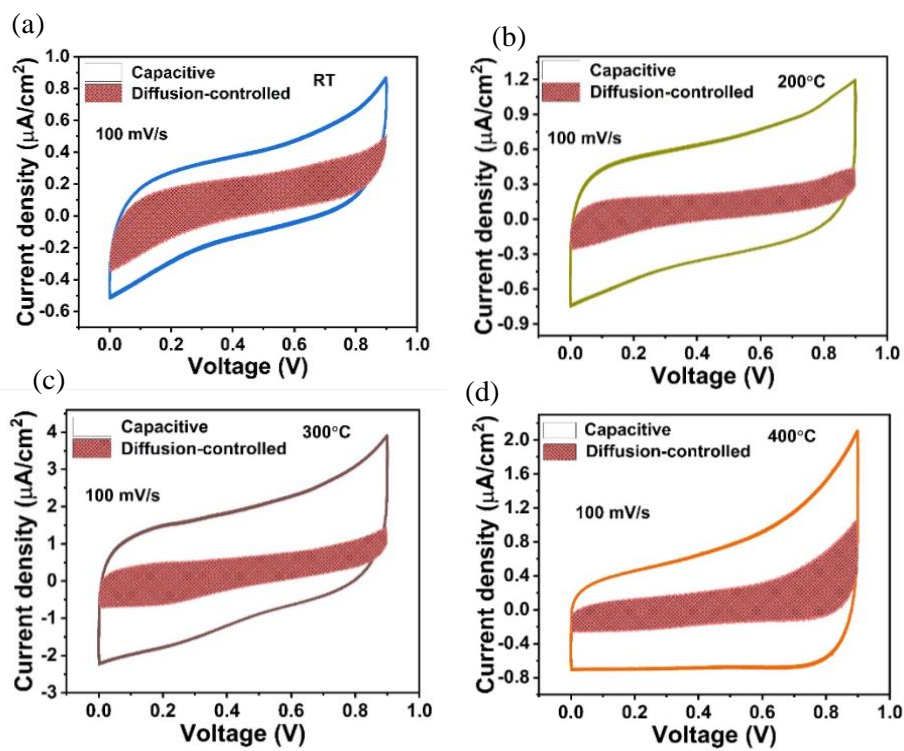

Figure 5: CV curve shows the capacitive and diffusion-controlled charge storage contributions at 100 mV/s scan rate in pristine Ti$_2$O$_3$ devices, (a) at RT, and annealed at (b) 200 °C, (c) 300 °C (d) 400 °C.

Figure 6(a)-(d) exhibit a typical separation of capacitive and diffusive contribution in pristine and annealed Ti$_2$O$_3$ devices at low scan rates of 5, 20, 50 and 100 mV/s. As shown in figure 6(a), the surface capacitive controlled process contributes 19%, 30%, 41% and 50% of the total charge at 5, 20, 50 and 100 mV/s, respectively for pristine Ti$_2$O$_3$ devices at RT, suggesting the dominant diffusion-controlled charge storage mechanism at lower scan rate. Although, there are no evident peaks in the CV responses, proton diffusion could contribute to the observed charge enhancement. Moreover, the surface capacitive contribution increases with increase of scan rate.

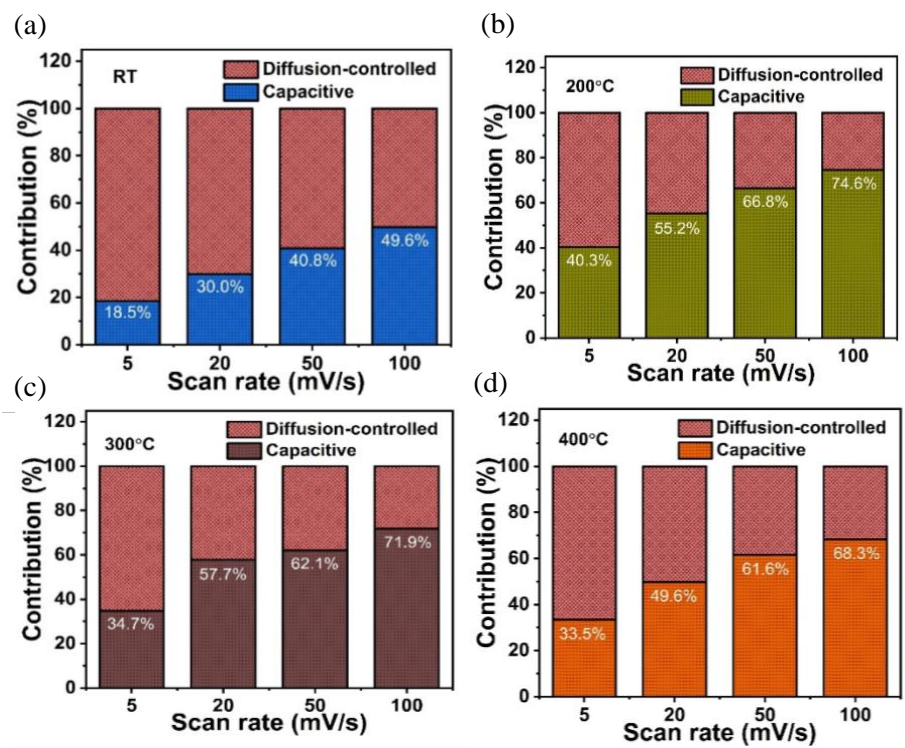

Figure 6: Capacitive and diffusion-controlled charge storage contribution at different scan rates of pristine Ti$_2$O$_3$ devices, (a) at RT, and annealed Ti$_2$O$_3$ devices at (b) 200 °C, (c) 300 °C (d) 400 °C.

It is noteworthy that upon annealing, the surface capacitive effect increases at all temperatures, resulting in significantly fast charge storage along with long term cyclic stability. Also, the increase in surface-capacitive contribution with increasing scan rate is observed after annealing of Ti$_2$O$_3$ devices and at higher scan rate of 100 mV/s, the charge storage mechanism is dominantly capacitive.

The areal capacitance or energy density are more suitable performance metrics for electrochemical response analysis compared to traditional gravimetric capacitance because the mass of active material is negligibly small compared with the mass of the whole device.[20] The areal capacitance of micro−electrochemical capacitors was calculated from the CV curves obtained at various scan rates by considering the area of both the interdigitated electrodes and the gap between adjacent fingers.

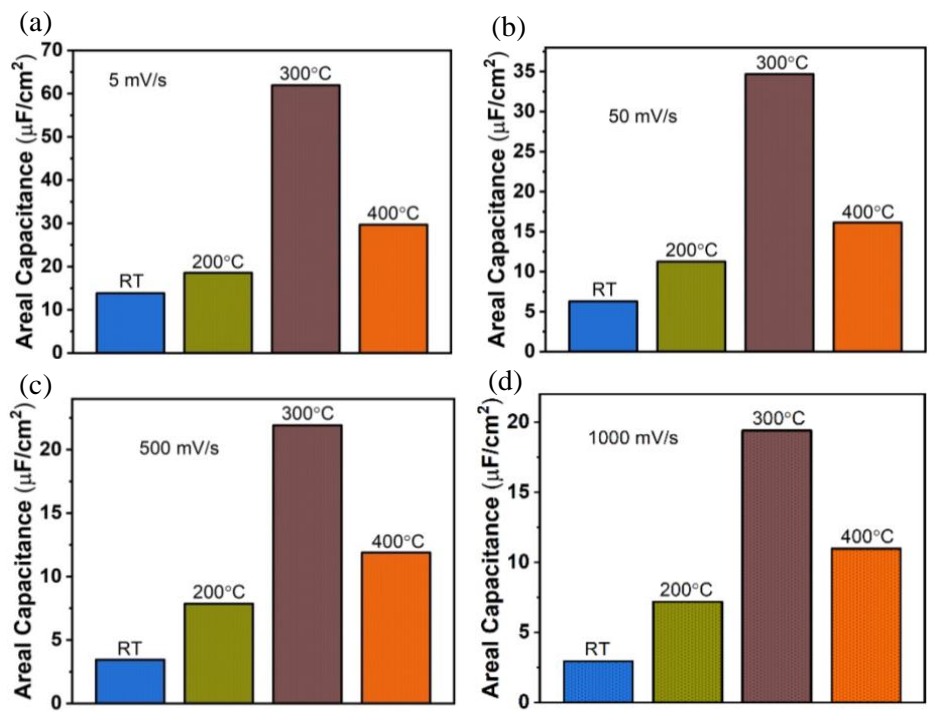

Figure 7. Areal electrochemical capacitance of pristine $Ti_2O_3$ and annealed at 200 °C, 300 °C and 400 °C at the scan rate of (a) 5 mV/s, (b) 50 mV/s, (c) 500 mV/s and 1000 mV/s.

Figure 7 summarises the calculated areal electrochemical capacitance at the four scan rates (5−1000 mV/s). As evident from figure 7(a−d) that at all the scan rates, the electrochemical capacitance is dominated by annealed $Ti_2O_3$ at 300 °C. The pristine $Ti_2O_3$ device showed an areal capacitance of about 13.84 µF/cm$^2$ at a scan rate of 5 mV/s, which reduced to 6.28 µF/cm$^2$ at a scan rate of 50 mV/s. Moreover, capacitance further dropped to 3.45 and 2.95 µF/cm$^2$ at the higher scan rates of 500 mV/s and 1000 mV/s, respectively. The capacitance at 500 mV/s and 1000 mV/s is about 24.9% and 21.3% of its initial capacitance at the scan rate of 5 mV/s. An enhancement of 34.24% in electrochemical capacitance is observed after annealing of granular $Ti_2O_3$ at 200 °C at a scan rate of 5 mV/s. The $Ti_2O_3$ device annealed at 300 °C showed an areal electrochemical capacitance of about 61.96 µF/cm$^2$ at a scan rate of 5 mV/s, which reduced to 29.96 µF/cm$^2$ at a scan rate of 100 mV/s. Moreover, capacitance dropped to 21.92 and 19.41 µF/cm$^2$ at the higher scan rates of 500 mV/s and 1000 mV/s, respectively. However, in $Ti_2O_3$ device after annealed at 300 °C, an anomalous increase of 559% is observed at a scan rate of 1000 mV/s and 347% after decreasing the scan rate to 5 mV/s compared with the pristine $Ti_2O_3$. An extremely large enhancement of 1153% in electrochemical capacitance was observed after ball−billing of the pristine granular $Ti_2O_3$ and annealed at 300 °C compared to pristine $Ti_2O_3$ device (without ball−billing) (figure S7(a)) and 180% enhancement when compared to pristine $Ti_2O_3$ device annealed at 300 °C (figure S7(b)). The enhancement is clearly attributed to the reduced particle size distribution providing higher accessible sites for the electrolyte (figure S7(c). The high specific surface area of ball−milled $Ti_2O_3$ (12.5 m$^2$/g) than pristine $Ti_2O_3$ (0.283 m$^2$/g) is confirmed by Brunauer−Emmett−Teller (BET) surface area measurement using nitrogen gas adsorption and desorption. The ball−milled $Ti_2O_3$ device exhibited a high capacitance of 173.5 µF/cm² at 5 mV/s, which reduced to 73.62 µF/cm² at 50 mV/s (figure S7(d)).

The electrochemical capacitor response of thermally stable phases of the $Ti_2O_3$ (Mott-insulator) after ex-situ annealing is evaluated and compared with the earlier reports; where annealing of graphene coated with $V_2O_5$, categorized as Mott−insulator, demonstrated a continuous decay in electrochemical capacitance to ~63% with the increase in the temperature till 500 °C.[21] Moreover, hydrous $RuO_2$ powder exhibits a sharp decay in the electrochemical capacitance when annealing temperature exceeded 150 °C, in a temperature range of 25 to 400 °C and the reported decrease in the specific capacitance was 2644%.[22]

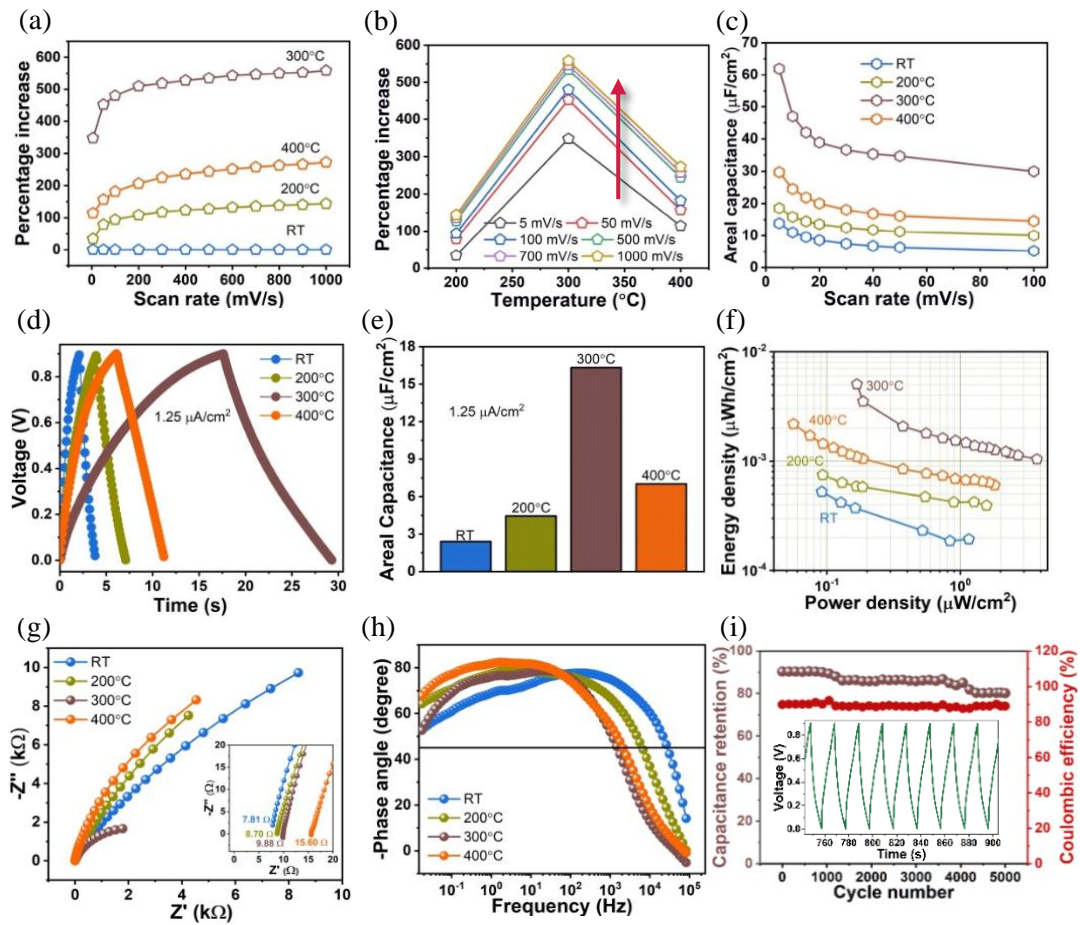

Figure 8. Electrochemical parameters of the $Ti_2O_3$ devices at different heating conditions: (a) Percentage increase in the electrochemical capacitance after annealing is measured with respect

to the pristine $Ti_2O_3$ at different scan rates. (b) Percentage increase in the electrochemical capacitance is measured with respect to the pristine response of the $Ti_2O_3$ at different annealing conditions. (c) Areal capacitance is shown at various scan rates. (d) Charge−discharge curves and (e) areal electrochemical capacitance at a current density of 1.25 µA/cm$^2$. (f) Ragone plot of $Ti_2O_3$ devices under various annealing conditions. (g) AC impedance spectroscopy of $Ti_2O_3$ devices; inset depicts a zoom−in view at high−frequency region. (h) Bode plot of the $Ti_2O_3$ devices. (i) Capacitance retention and Coulombic efficiency of $Ti_2O_3$ device annealed at 300 °C plotted for 5000 cycles; inset depicts a few charge-discharge cycles.

Figure 8(a) depicts the percentage enhancement in electrochemical capacitance from RT to annealed $Ti_2O_3$ devices at 200 °C, 300 °C and 400 °C at the scan rates ranging from 5 to 1000 mV/s. Although the device annealed at 400 °C has higher electrochemical capacitance than the device annealed at 200 °C, but it remains lower compared to the device annealed at 300 °C at all the scan rates. Moreover, the electrochemical capacitance of $Ti_2O_3$ after annealing is enhanced compared to the pristine device at all the scan rates. A faster increase in the electrochemical capacitance is observed at the lower scan rates, while all the devices showed a slow increase at higher scan rates. Figure 8(b) clearly depicts a sharp increase in the percentage enhancement of electrochemical capacitance of $Ti_2O_3$ annealed at 300 °C at all the scan rates. Further, the areal electrochemical capacitance is shown to decrease with the scan rate measured in the range of 5−100 mV/s (figure 8(c)) in all the $Ti_2O_3$ devices.

The performance of $Ti_2O_3$ devices was further evaluated by measuring the galvanostatic charge−discharge (CD) curve at various applied current densities. The CD curves show a triangular shape, which is a typical EDLC behaviour at a current density of 1.25 µA/cm$^2$ (figure 8(d)). A sudden drop in the voltage (IR drop) at the beginning of the discharge curves is a measure of the overall resistance of the device.[23] A small IR drop in all the devices after annealing indicates a very low internal resistance and the values are given by 0.0271, 0.0103 and 0.0094 V for the devices annealed at 200 °C, 300 °C, and 400 °C, respectively (figure S8(a)). The areal specific capacitances were also calculated from the CD curves and plotted as shown in figure S8(b). A higher electrochemical capacitance at all the current densities was achieved for the $Ti_2O_3$ device annealed at 300 °C, which is in good agreement with CV results. The pristine $Ti_2O_3$ device showed a maximum areal capacitance of 9.52 µF/cm² at a current density of 0.167 µA/cm². The maximum areal capacitance of annealed $Ti_2O_3$ device (300 °C) is measured 45.2 µF/cm² at a current density of 0.37 µA/cm², which reduced to 10.21 µF/cm² at much increased current density of 8.75 µA/cm². Electrochemical capacitance from the CD measurement of $Ti_2O_3$ devices is compared in figure 8(e), which shows an enhancement of 580% for the $Ti_2O_3$ device annealed at 300 °C from pristine $Ti_2O_3$ device at a current density of 1.25 µA/cm². Therefore, CD measurements further confirm the annealing effect on the electrochemical capacitance of granular $Ti_2O_3$, where the electrochemical capacitance of the sample annealed at 400 °C sharply dropped by 132% from the sample annealed at 300 °C. Further enhancement in electrochemical capacitance was observed in ball−milled $Ti_2O_3$, where the device annealed at 300 °C exhibited an areal capacitance of 183.37 µF/cm² at 1.04 µA/cm² current density, which reduced to 13.57 µF/cm² at a current density of 20.84 µA/cm² (figure S8(c)). A further enhancement of 584% is noted in annealed ball−milled $Ti_2O_3$ device compared to annealed $Ti_2O_3$ device at a current density of 1.25 µA/cm², when both the devices are annealed at 300°C (figure S8(d)).

Areal energy density and power density of all the four devices, including pristine $Ti_2O_3$, were calculated from the CD curve at various current densities and shown in the Ragone plot (figure 8(f)). The measured drop in energy density with increasing power density is small, as evident in the plot. The pristine $Ti_2O_3$ device demonstrated an areal energy density of 1.034 nWh/cm² at a power density of 73.66 nW/cm² and 0.193 nWh/cm² at 1159.3 nW/cm². The Ragone plot of $Ti_2O_3$ device annealed at 300 °C shows substantially higher energy density compared to the other three devices. For instance, the $Ti_2O_3$ device (annealed at 300 °C) achieved a maximum energy density of 5.04 nWh/cm$^2$ at a power density of 168 nW/cm$^2$ and a maximum power density of 3749.3 nW/cm$^2$ is noted at the energy density of 1.041 nWh/cm$^2$. For quantitative analysis, the areal energy density is plotted in a bar diagram at a current density of 1.25 µA/cm$^2$ for all the four $Ti_2O_3$ devices (figure S9(a)). The $Ti_2O_3$ device annealed at 300 °C exhibited 675.75% enhancement in areal energy density compared to pristine one (figure S9(b)). Furthermore, an enhancement of 549% in the areal energy density was observed in ball−milled $Ti_2O_3$ device, compared to without ball−milled and annealed at 300 °C (figure S9(c)). The ball−milled $Ti_2O_3$ device annealed at 300 °C attains a high energy density of 19.23 nWh/cm$^2$ at a power density of 452.6 nW/cm$^2$ and a maximum power density of 8010.4 nW/cm$^2$ at the energy density of 1.11 nWh/cm$^2$ (figure S9(d)).

Electrochemical impedance spectroscopy was performed in terms of Nyquist and Bode plots to further provide an incite in the ohmic and diffusion resistances of ions in the electrodes for all the four devices (figures 8(g) and (h)). The equivalent series resistance values of pristine and annealed $Ti_2O_3$ devices, as shown in the inset of figure 8(g) are 7.81, 8.70, 9.88, and 15.60 Ω thus, showing a good electrochemical conductivity of all the devices. A semi−circular arc is observed only at 300 °C, which explains the space charge polarization. The dependence of phase angle with frequency (Bode plot) in the $Ti_2O_3$ devices was further presented in figure 8(h) for informative analysis of impedance spectroscopy. For frequencies up to 100 Hz, all the phase angles of $Ti_2O_3$ devices are close to 90°; after that, the phase angle decreases with an increased frequency. In addition, the characteristics frequency ($f_0$) at a phase angle of −45° indicates that the resistive and capacitive impedances are equal and later, resistive behaviour dominates.[23] As shown in figure 8(h), the characteristics frequency of pristine $Ti_2O_3$ is 27711.35 Hz, and 6463.05, 1505.42 and 2136.5 Hz for $Ti_2O_3$ devices annealed at 200 °C, 300

°C, and 400 °C, respectively. Therefore, the corresponding relaxation time constant $\tau_0 (= 1/f_0)$, defined as the minimum time required to discharge all the energy from the device with an efficiency greater than 50%,[24,25] is 0.036 ms for pristine $Ti_2O_3$, in contrast to 0.154, 0.664, and 0.468 ms for $Ti_2O_3$ devices annealed at 200 °C, 300 °C and 400 °C, respectively. The obtained time constant for $Ti_2O_3$ devices is promising compared to the previously best−reported value for the EDLC based micro electrochemical capacitors, such as activated carbon (700 ms),[24] onion like carbon (26 ms),[24] laser reduction graphene (19 ms)[26], and graphene/CNT composite (3.4 ms).[23] Furthermore, the cyclic life of $Ti_2O_3$ device annealed at 300 °C was evaluated by long term charging and discharging cycle over 5000 cycles at a current density of 0.167 µA/cm² (figure 8(i)). A high Coulombic efficiency of 89% is measured, while, capacitance retention was maintained 80% after 5000 cycles.

Semiconductor $Ti_2O_3$ with a bandgap of 0.11 eV further allows radiation interaction in the IR range. The electrochemical measurements were performed in the presence of IR radiation using a bulb (Philips Infrared−R95E). The light interaction further improved the electrochemical capacitance of both pristine and annealed $Ti_2O_3$ devices. Figure 9(a) shows the CV measurement of pristine $Ti_2O_3$ device at three conditions of before and after IR illumination for 30 minutes and the measurement conducted after 24 h of illumination. Figure 9(b) summarises the enhanced electrochemical capacitance of all the $Ti_2O_3$ devices compared to without IR illumination.

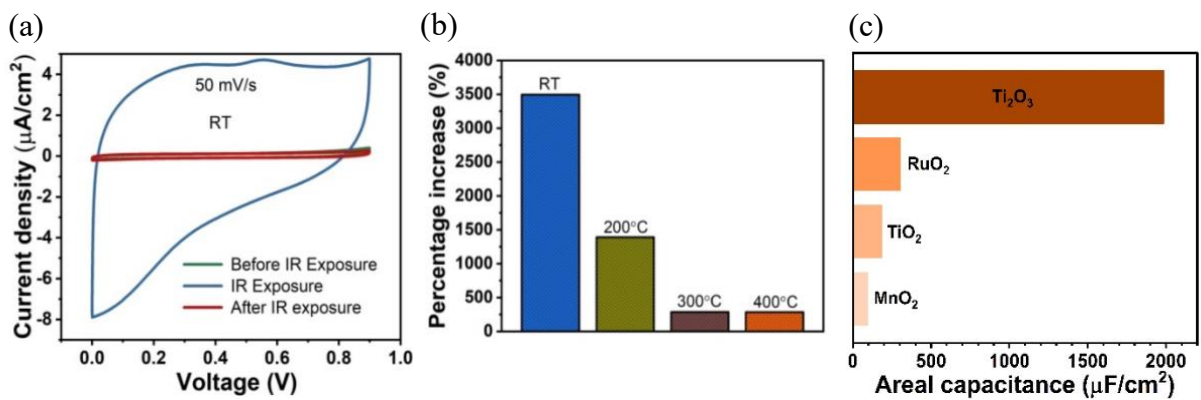

Figure 9. (a) CV curves of pristine Ti$_2$O$_3$ device before, in presence of IR and after IR interaction. (b) The percentage enhancement in electrochemical capacitance upon IR interaction is measured at various thermal conditions compared to without illumination condition. (c) Areal capacitance of ball−milled Ti$_2$O$_3$ electrode (annealed at 300 °C) using three-electrode configuration in 1M H$_3$PO$_4$ compared with previously reported literature.

Under IR illumination, an enormously large enhancement of 3494% is observed in the pristine Ti$_2$O$_3$ device at a scan rate of 50 mV/s. The effect of metallic transition with heating is clearly confirmed with the decay in the electrochemical response of the annealed Ti$_2$O$_3$ devices upon IR illumination. A much larger enhancement in electrochemical capacitance upon IR interaction does not have any impact on electrochemical characteristics of the device as shown in figure S10, where all the samples exhibited reversible behaviour after turning off the IR illumination.

Furthermore, the electrochemical performance of ball−milled Ti$_2$O$_3$ device annealed at 300 °C is also characterized in aqueous H$_3$PO$_4$ electrolyte. The measured areal capacitances from the CV curves are 1052.7 and 706.5 µF/cm$^2$ at a scan rate of 1 and 5 mV/s, respectively. Also, the electrochemical performance of ball−milled Ti$_2$O$_3$ material (annealed at 300 °C) analysed using a three-electrode configuration in 1M H$_3$PO$_4$ electrolyte (details are in supporting information). The measured capacitance from the CV curve is 2409 mF/g (1985 µF/cm$^2$) at a scan rate of 1 mV/s as shown in figure S11. The as-obtained capacitance is compared with previously reported literature values (figure 9(c)) e.g., TiO$_2$/powder (33−181) µF/cm$^2$ at 100 to 1 mV/s,[27] double-layer capacitance from colloidal RuO$_2$ (300 µF/cm$^2$),[28] three-dimensional porous MnO$_2$ structure (70-90 µF/cm$^2$ ).[29]

Based on both experimental and simulation results, the anomalous electrochemical enhancement of Ti$_2$O$_3$ under the annealed and IR illumination conditions can be explained by its bandgap and the subsequent bandgap crossover of Ti 3$d$ $a_{g1}$ and $e_g^\pi$ levels as explained by Goodenough.[30] Figure 10(a) schematically depicts the three−dimensional network of Ti$_2$O$_3$ coated on the electrode plane for the symmetric electrochemical capacitor.[31] The crystal structure of Ti$_2$O$_3$ mainly consists of honeycomb lattice in parallel layers formed by Ti$^{3+}$ ions in the $a$–$b$ plane, which are paired along the $c$−axis as dimers and each ion is surrounded by the O$^{2-}$ ions.[32] The response of pristine semiconductor Ti$_2$O$_3$ toward electrochemical capacitance is limited by the space−charge region as depicted by figure 10(b) near its surface.[1] The band overlapping at higher annealing temperature changes the density and distribution of the resulting delocalized electrons leading to a jellylike structureless distribution (schematic in figure 10(c)).[1] The rise in temperature from RT to 300 °C and an enhanced EDLC response clearly indicated a large charge accumulation on the surface with more DOS, as explained earlier in figure 3.

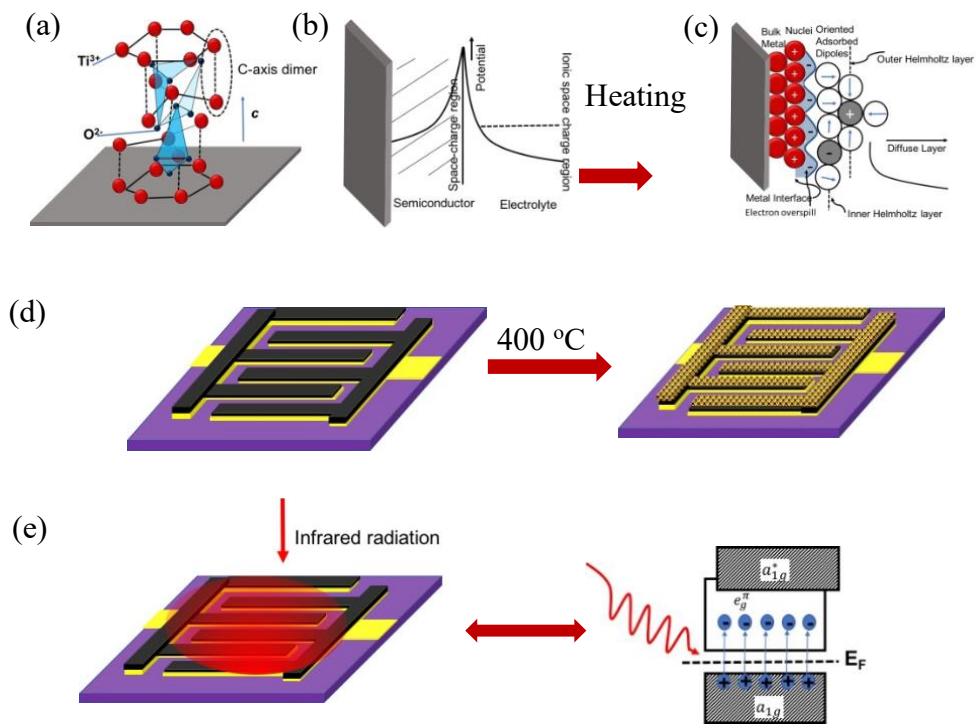

Figure 10. (a) Schematic representation of crystal structure of $Ti_2O_3$. (b) Space charge distribution of charge carriers within the pristine semiconducting $Ti_2O_3$ near the electrode/electrolyte interface. (c) Charge layer formation at the interface of metallic $Ti_2O_3$ and electrolyte (both the figures (b) and (c) are adapted from reference 1). (d) Schematic diagram to depict the surface transformation of $Ti_2O_3$ into $TiO_2$ at 400 °C in the device. (e) IR interaction of the electrochemical capacitor (left) and excitons generation across the bandgap of pristine $Ti_2O_3$.

As mentioned in our simulation results that the structural disorder in Ti−Ti dimers along the $c$−axis allowing Ti ions hopping to the $a−b$ plane *via* $e_g^{\pi}$ channel possibly attributes to the anomaly in electrochemical capacitance response.[32] Although we have not observed any significant changes in the lattice parameters in the XRD measurements, these localized distortions of the dimers are reported by Hwang *et al.*[33] through the X−ray absorption fine structure measurements. Therefore, sharing the *3d* electrons between the $c$−axis and basal−plane orbitals enhances the polarizability following the charge hopping[34] led to an anomalous enhancement in electrochemical capacitance at 300 °C. Further annealing at 400 °C has shown a clear signature of appearance of $TiO_2$ phase and coexistence of both $Ti_2O_3$ and $TiO_2$ as described earlier and schematically depicted in figure 10(d) by following the phase transformation from $Ti_2O_3 \rightarrow TiO_{2-x} \rightarrow TiO_2$ as given by Li *et al.*[5] The IR interaction with pristine and annealed $Ti_2O_3$ can be understood through the additionally enhanced space−charge contribution from the transition across the bandgap as shown in figure 10(e). The IR induced enhanced electrochemical capacitance in annealed $Ti_2O_3$ after transition to a metallic state could be attributed to the thermal excitations through the vibration of crystal lattice.

**Conclusion**

The insulator to metal transition in Mott−insulator, $Ti_2O_3$ is exploited for the electrochemical capacitance. An extraordinary large enhancement in electrochemical capacitance of ~560% is observed after annealing the sample at 300 °C compared to the other annealing conditions at 200 °C and 400 °C. Further pristine $Ti_2O_3$ provided an enhancement of ~3500% in electrochemical capacitance upon infrared radiation interaction. These extraordinary enhancements in the electrochemical responses are explained based on the interaction of Ti ions with the O ions upon heating conditions and resulting localized hybridization leading to the electron spill over the surface of the metallic $Ti_2O_3$. The DFT simulations further supported the enhanced participation of surface states with the increase in temperature, which strongly supports our experimental findings.